\documentclass[lettersize,conference]{IEEEtran}
\usepackage{cite}
\usepackage{amsmath,amssymb,amsfonts}
\usepackage{algorithmicx}
\usepackage{graphicx}
\usepackage{textcomp}
\usepackage[nolist]{acronym}
\usepackage{longtable}
\usepackage{relsize}
\usepackage{threeparttable}
\usepackage[normalem]{ulem}
\usepackage{comment}
\usepackage{color,soul}
\usepackage{multicol,multirow,xcolor,colortbl}
\usepackage{makecell}
\usepackage{orcidlink}
\usepackage{xparse}
\usepackage{caption}
\usepackage{subcaption}
\usepackage{algpseudocode}
\usepackage{svg}

\usepackage{xcolor}
\usepackage{etoolbox}

\begin{acronym} 
\acro{ADAS}{advanced driver assistance systems}
\acro{ADC}{analog to digital conversion}
\acro{CA}{collision avoidance}
\acro{CA-CFAR}{cell-averaging constant false alarm rate}
\acro{CbC}{chirp-by-chirp}
\acro{CD}{collision detection}
\acro{CDF}{cumulative distribution function}
\acro{EIRP}{effective isotropic radiated power}
\acro{FbF}{frame-by-frame}
\acro{FbF-CA}{frame-by-frame collision avoidance}
\acro{FbF-CA-FBS}{frame-by-frame collision avoidance with full-band sensing}
\acro{FbF-CD}{frame-by-frame collision detection}
\acro{FBS}{full-band sensing}
\acro{FOV}{field of view}
\acro{FMCW}{frequency modulated continuous wave}
\acro{INR}{interference to noise ratio}
\acro{LOS}{line of sight}
\acro{NLOS}{non line of sight}
\acro{OFDM}{orthogonal frequency division multiplexing}
\acro{RCS}{radar cross-section}
\acro{SINR}{signal to interference and noise ratio}
\acro{SW}{software simulation}
\end{acronym}

\newcommand{\BW}{{B_\text{ch}}}

\newcommand{\BTOT}{{B_\text{TOT}}}
\newcommand{\fstartTOT}{{f_\text{start}}}
\newcommand{\fstopTOT}{{f_\text{stop}}}
\newcommand{\duty}{{\delta}}

\newcommand{\Tchirprep}{{T_\text{rch}}}
\newcommand{\Tchirp}{{T_\text{ch}}}
\newcommand{\Mf}{{M_\text{f}}}

\newcommand{\BADC}{{B_\text{ADC}}}

\newcommand{\kch}{{K_\text{ch}}}
\newcommand{\nch}{{N_\text{ch}}}

\newcommand{\Tfail}{{T_\text{fail}}}
\newcommand{\Tframe}{{T_\text{rf}}}

\newcommand{\dmax}{d_\text{max}}
\newcommand{\RCS}{\sigma}

\newcommand{\INRmin}{\gamma_\text{min}}

\newcommand{\fzero}{f_0}
\newcommand{\EIRPdBm}{\epsilon_\text{dBm}}
\newcommand{\Gr}{{G_\text{r}}_\text{dB}}
\newcommand{\Boltzmann}{k_\text{B}}
\newcommand{\Tzero}{T_0}
\newcommand{\NF}{{F_\text{N}}_\text{dB}}
\newcommand{\NFlin}{{F_\text{N}}}
\newcommand{\Prmin}{{P_{\text{r}_\text{min}}}}
\newcommand{\deq}{d_\text{eq}}
\newcommand{\xchirp}{{x_\text{chirp}}}

\begin{document}

\title{Collision Avoidance and Collision Detection Strategies to Counteract Large-Scale FMCW Automotive Radar Interference}

\author{
\IEEEauthorblockN{
Francesco Miccoli\IEEEauthorrefmark{1},
Vittorio Todisco\IEEEauthorrefmark{1}\IEEEauthorrefmark{2},
Fabrizio Cuccoli\IEEEauthorrefmark{3},\\Matteo Passeri\IEEEauthorrefmark{3}, Vincent Martinez\IEEEauthorrefmark{4}, and
Alessandro Bazzi\IEEEauthorrefmark{1}\IEEEauthorrefmark{2} }\\
\IEEEauthorblockA{%
  \IEEEauthorrefmark{1}National Laboratory of Wireless Communications (WiLab), CNIT, Bologna, Italy\\
\IEEEauthorrefmark{2}DEI, University of Bologna, Italy\\
  \IEEEauthorrefmark{3}National Laboratory of Radar and Surveillance Systems (RaSS), CNIT, Pisa, Italy\\
  \IEEEauthorrefmark{4}NXP Semiconductors, Toulouse, France\\
}%
\thanks{Corresponding author: F. Miccoli (email: francesco.miccoli@wilab.cnit.it).}}

\maketitle

\begin{abstract}
Automotive radars operating in dense traffic environments may experience significant mutual interference, potentially degrading sensing reliability for advanced driver assistance systems (ADAS). This issue motivates the development of techniques to mitigate interference among multiple radar-equipped vehicles sharing the same spectrum.
This paper investigates interference
mitigation strategies for uncoordinated radar networks through system-level simulation. Two approaches are considered: frame-by-frame collision avoidance (FbF-CA), which exploits idle intervals to measure the perceived interference and possibly adapt the operating frequency, and frame-by-frame collision detection (FbF-CD), which estimates interference during transmission based on observed chirp collisions. An event-driven radar network simulator is used to model realistic highway traffic and radar interactions.
Results show that collision avoidance provides higher robustness when a larger bandwidth is available, while collision detection is more effective under bandwidth-limited conditions. The analysis highlights the trade-off between collision avoidance and collision detection mitigation strategies and provides design insights for scalable automotive radar deployments.%

\end{abstract}

\section{Introduction}
Automotive radars represent a key sensing technology for \ac{ADAS} and automated driving. Compared to other sensing modalities such as cameras or lidar, radar sensors provide reliable detection and tracking capabilities even under adverse weather conditions and at long distances. For these reasons, modern vehicles are increasingly equipped with multiple \ac{FMCW}-based radar sensors operating in the millimeter-wave bands in the 76--81~GHz band, with regional variations.
Thanks to their robustness and relatively low cost, radar sensors are expected to play an important role in future intelligent transportation systems.

\medskip
However, the growing penetration of radar-equipped vehicles also introduces new challenges. Radar signals transmitted by nearby vehicles may overlap in time and frequency, generating interference that can affect sensing reliability. In \ac{FMCW} automotive radars, such interference can produce signal components that resemble legitimate reflections, potentially leading to corrupted chirps, degraded sensing performance, or even the detection of false targets. This problem is further exacerbated by the fact that automotive radars typically operate in a shared spectrum with limited coordination among devices. As radar adoption continues to increase, the probability that multiple sensors operate simultaneously within the same spectrum increases, raising concerns about the scalability of current radar spectrum allocations in dense traffic scenarios \cite{8828037}.

Several studies have investigated the impact of radar-to-radar interference in vehicular environments, highlighting that the currently available spectrum may become insufficient in dense traffic scenarios \cite{SAE_Meeting_Report,https://doi.org/10.1049/rsn2.12096,11241101,11422277}. %
While on the one hand the use of additional spectrum around 140~GHz is being investigated~\cite{10734940}, on the other interference mitigation methods are also being considered. %
Interference mitigation can be achieved either by reducing it a priori, before transmission (proactive approaches) or by processing the received signal after interference occurs (reactive approaches). While reactive approaches can be used in addition to proactive ones, they are outside the scope of this work.

Focusing on proactive approaches, they can be broadly categorized into \emph{coordination-based} and \emph{uncoordinated} techniques. %
Coordination-based solutions rely on explicit cooperation among radar devices but require communication and standardization across different manufacturers, which may limit their practical deployment \cite{roudiere2021first}. Uncoordinated approaches instead rely on local strategies such as randomization of transmission parameters or sensing-based adaptation of the operating frequency. Examples include frame-by-frame or chirp-by-chirp frequency hopping techniques and random-access %
radar transmission schemes \cite{Jin2020,Jin2021,Haritha2021,k2022slotted}. 

Very few works have included mechanisms that perform measurements to deal with interference \cite{Kurosawa2019,ishikawa2019,k2022slotted}. All of these works consider only the sensed power and introduce actions in the time domain.
In \cite{Kurosawa2019,ishikawa2019}, the authors propose to add a carrier sensing mechanism before each transmission, with random backoff in the case a signal is received. The proposal is investigated assuming no propagation delay, which means an ideal and unrealistic scenario where radars are co-located. A similar idea is explored in \cite{k2022slotted}, where time is slotted and sensing is performed in the slots, possibly followed by random backoff. All results appear preliminary and are obtained with one or a few interferers, without considering system-level interactions among multiple radar-equipped vehicles operating in realistic traffic scenarios.

In this paper, we investigate mitigation strategies based on measurements for automotive radar interference through system-level simulations in realistic vehicular environments. In particular, we consider two complementary approaches: a \ac{CA} mechanism that adapts the operating frequency based on interference measurements during idle periods and a \ac{CD} mechanism that updates the operating frequency based on repeated chirp collisions observed during radar operation.
An event-driven radar network simulator is used to model radar transmissions and interference interactions in dense highway scenarios with multiple vehicles.

\medskip
The main contributions of this work are:
\smallskip
\begin{itemize}
\item The development of a system-level simulation framework for analyzing radar interference in realistic vehicular traffic scenarios.
\item The definition and implementation of sensing-based mitigation strategies based on \ac{CA} and \ac{CD}.
\item A comparative evaluation of the considered strategies.%
\end{itemize}

\section{System Modeling}\label{sec:WiLabVIsim}

Modeling a highway scenario with many interfering radar-equipped vehicles requires balancing accuracy and computational complexity. In this work, we use the approach detailed in \cite{11422277} and hereafter recalled. 

\begin{figure}[t]
\centering
\subfloat[]{%
\centering
         \includegraphics[width=0.9\columnwidth]{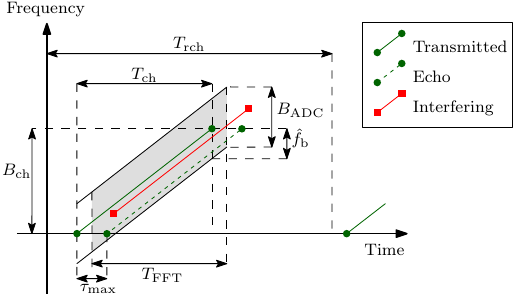}
         \label{fig:chirp}    }
\\
\subfloat[]{%
\centering
         \includegraphics[width=0.88\columnwidth]{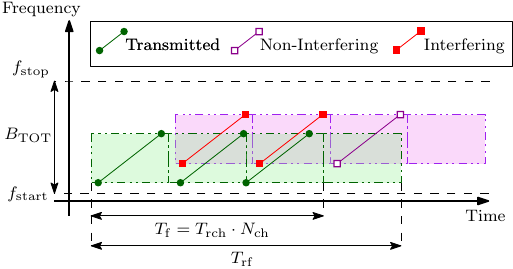}
         \label{fig:frame}    }
\caption{Representation of (a) the chirp and (b) the frame.
}\label{fig:frameandchirp}
\vspace{-\baselineskip}
\end{figure}

\subsection{FMCW Radar Signal Model}\label{subsec:FMCWradar}

\ac{FMCW} radars operate by transmitting a sequence of chirps, where each chirp is a signal whose frequency increases linearly over time within a given bandwidth. The transmitted chirps are reflected by surrounding objects, generating delayed echoes that are processed to estimate target range and velocity.
The structure of a chirp is shown in Fig.~\ref{fig:chirp}. Each chirp spans a bandwidth $\BW$ and has duration $\Tchirp$. Chirps are transmitted periodically within a repetition interval $\Tchirprep$, which includes $\Tchirp$ and an idle time required to reset the oscillator. The starting instant of each chirp is assumed to vary randomly within $\Tchirprep$, from chirp to chirp.
A frame consists of $\nch$ consecutive chirps, as illustrated in Fig.~\ref{fig:frame}, followed by an additional idle interval. The total frame duration, including active and inactive periods, is denoted as $\Tframe$. The duty cycle $\duty$ is defined as the ratio between active transmission time and total frame duration, i.e., \mbox{$\duty = \Tchirprep \cdot \nch / \Tframe$}.

\subsection{Interference Modeling and Failure Abstraction}\label{Sec:Modeling}

In \ac{FMCW} radars, interference arises when signals transmitted by neighboring radar sensors overlap in time and frequency with the chirps of another radar. Depending on the characteristics of the interfering signals, a distinction is commonly made between \emph{uncorrelated} and \emph{correlated} interference. Uncorrelated interference results from different \ac{FMCW} waveforms, leading to an increase in the noise floor, whereas correlated interference occurs when interfering radars use similar chirp patterns, potentially generating ghost targets.

In this work, all radars are assumed to operate within the same frequency band, spanning from $\fstartTOT$ to $\fstopTOT=\fstartTOT+\BTOT$, and to use identical waveform parameters, resulting in correlated interference. Under this assumption, interfering chirps have the same slope and may overlap, as illustrated in Fig.~\ref{fig:frameandchirp}.
This assumption will be relaxed in future work; however, in practice, radars designed for different functions (e.g., long-range front radars and medium-range corner radars) typically use different bands.

Interference is modeled at the chirp level using the chirp and frame structure described in Section~\ref{subsec:FMCWradar}, based on time-frequency overlap conditions. In particular, a chirp is affected only if an interfering signal falls within the \ac{ADC} observation bandwidth during its reception interval.
Following the approach in \cite{11422277}, a chirp is considered corrupted if such an overlap persists for at least a fraction $\xchirp$ of the chirp duration. This corresponds to severe interference, potentially leading to chirp loss within a frame.
A frame is considered lost if the number of corrupted chirps is greater than or equal to $\kch$. Finally, a system failure is declared when $\Mf$ consecutive frames are lost.
The symbols used in the model are summarized in Table~\ref{tab:settings}.

\subsection{Interferer Selection and Propagation Model}%

Given the potentially large number of radars in the scenario, considering all mutual interactions would be computationally prohibitive and often unnecessary. We therefore restrict the analysis to \emph{potential interferers}, defined as those radars whose signals fall within the \ac{FOV} of a receiver with sufficient power to produce non-negligible interference. %

At each time instant, potential interferers are identified from the positions of vehicles and their associated radars by considering both direct and first-order reflected propagation paths,\footnote{At millimeter-wave frequencies%
, higher-order reflections are typically negligible due to severe propagation losses.} and retaining only those sources for which the resulting \ac{INR} exceeds a minimum threshold $\INRmin$. For each interferer, only the dominant propagation path is considered, corresponding to the direct path under \ac{LOS} conditions or the least attenuated reflected path under \ac{NLOS} conditions.

In the case of \ac{LOS}, denoting by $P_\text{t}$ the transmission power, $G_\text{t}$ and $G_\text{r}$ the transmitting and receiving antenna gains, respectively, $\lambda$ the signal wavelength, and $d$ the distance between transmitter and receiver, the received power $P_{\text{r}_\text{LOS}}$ can be calculated as 
\begin{equation}\label{eq:A0}
    P_{\text{r}_\text{LOS}} = \frac{P_\text{t}G_\text{t}G_\text{r}\lambda^2}{(4 \pi d)^2}\;.
\end{equation}

Denoting by $\Prmin$ the minimum received power required to exceed $\INRmin$, the maximum distance $\dmax$ is given by
\begin{equation}\label{eq:A1}
    \dmax = \sqrt{\frac{P_\text{t}G_\text{t}G_\text{r}\lambda^2}{\Prmin (4 \pi)^2}}\;.
\end{equation}
where
\begin{equation}\label{eq:prmin}
    \Prmin = \INRmin \cdot \Boltzmann \cdot \Tzero \cdot \BADC \cdot \NFlin
\end{equation}
with $\Boltzmann$ denoting the Boltzmann constant, $\Tzero$ the reference temperature of 290~K, $\BADC$ the \ac{ADC} bandwidth, and $\NFlin$ the receiver noise figure.

In the case of reflected interference (i.e., NLOS), the signal propagates over two paths, denoted as $d_1$ and $d_2$, corresponding to the distances from the interferer to the reflection point and from the reflection point to the receiver, respectively. Using the same definitions as for the LOS case and introducing the \ac{RCS} $\RCS$, the received power $P_{\text{r}_\text{NLOS}}$ is given by
\begin{equation}\label{eq:A2}
    P_{\text{r}_\text{NLOS}} = \frac{P_\text{t}G_\text{t}G_\text{r}\lambda^2}{(4 \pi)^2 d_1^2}\;\frac{\RCS}{4 \pi d_2^2}\;.
\end{equation}

From \eqref{eq:A0} and \eqref{eq:A2}, an equivalent distance $\deq$ can be derived as
\begin{equation}\label{eq:disteq}
    \deq = \sqrt{\frac{4 \pi d_1^2 d_2^2}{\RCS}}
\end{equation}
which corresponds to the distance at which an interfering source in LOS would produce the same received power. A reflected interferer is therefore considered a potential interferer if and only if $\deq \leq \dmax$.

\section{Sensing-Based Interference Mitigation Proposals}

Hereafter, the \ac{FbF-CA} and the \ac{FbF-CD} proposals are described.

\subsection{Frame-by-Frame Collision Avoidance}

In the \ac{FbF-CA} method, each radar starts by randomly selecting a starting frequency between $\fstartTOT$ and $\fstopTOT - \BW$. Then, the same carrier frequency is maintained unless a certain level of interference power is measured during the idle interval %
between two consecutive frame transmissions.

\begin{figure}[t]
    \centering
    \includegraphics[width=0.7\columnwidth]{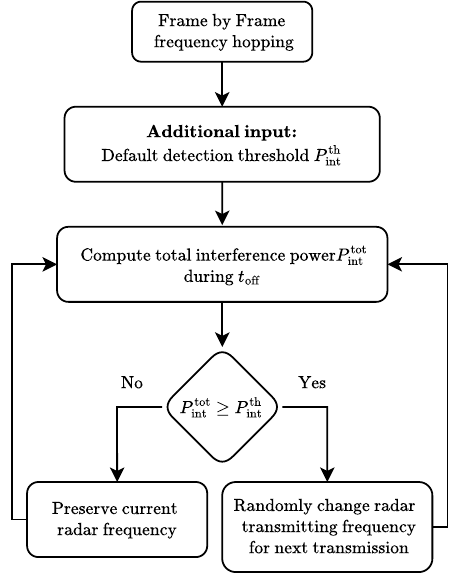}
    \caption{Frame-by-frame collision avoidance flow diagram.}%
    \label{fig:fbf_ca_flow}
\end{figure}

More specifically, as shown in the flow diagram %
in Fig.~\ref{fig:fbf_ca_flow},
the core step of this technique is the computation of the total detected interference power received from all the interfering radars during the non-transmitting period $t_\text{off}$, called $P_\text{int}^\text{tot}$. 
Assuming that there are $I$ potential interferers, the interference power contribution from the $i$-th potential interferer is given by
\begin{equation}
    P_{{\text{int}}_i} = P_{{\text{chirp}}_i} \cdot \eta_i \cdot N_i,
\end{equation}
where $P_{{\text{chirp}}_i}$ is the interference power per chirp, $\eta_i$ denotes the fraction of the interfering frame in the frequency domain that overlaps with the sensing window, and $N_i$ represents the number of chirps that fall within the sensing window. The total detected interference power is then obtained by summing the contribution of all interferers as
\begin{equation}
    P_\text{int}^\text{tot} = \sum_{i=1}^{I} P_{{\text{int}}_i}.
\end{equation}

If the detected interference $P_\text{int}^\text{tot}$ exceeds a predefined threshold $P_\text{int}^\text{th}$, i.e., if $P_\text{int}^\text{tot}\geq P_\text{int}^\text{th}$, a new random starting frequency is selected between $\fstartTOT$ and $\fstopTOT - \BW$ for the transmission of the next frame. %

\begin{figure}[t]
    \centering
    \includegraphics[width=0.85\columnwidth]{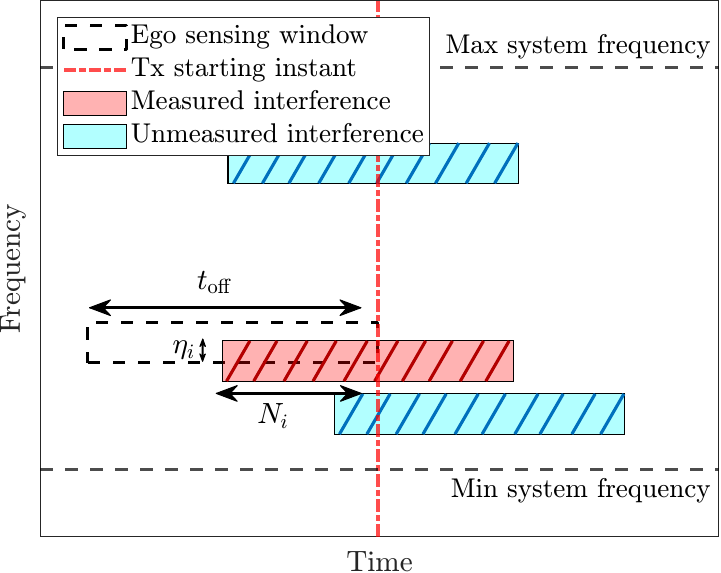}
    \caption{Example of interference power detection with frame-by-frame collision avoidance. 
    \newline}
    \vspace{-5mm}
    \label{fig:fbf_ca_sensing}
\end{figure}

Figure \ref{fig:fbf_ca_sensing} %
portrays the proposed method from %
the perspective of a generic radar suffering from interference.
In the figure, the inactive interval after the frame transmission is represented by a rectangle of time duration $t_\text{off}$ and bandwidth $\BW$, also referred to as the sensing window. At the end of this interval, the given radar performs the subsequent frame transmission based on the interference detected inside the sensing window. The red and light blue rectangles in Fig.~\ref{fig:fbf_ca_sensing} represent the frames transmitted by potential interferers that do and do not occupy the same time-frequency resources within the sensing window, respectively.

\subsection{Frame-by-Frame Collision Detection}

In the \ac{FbF-CD} method, again each radar starts by randomly selecting a carrier frequency within  $\fstartTOT$ and $\fstopTOT - \BW$ for its transmissions. Then, the starting frequency for the next frame transmission is changed if a certain interference is measured during the transmission. To identify the presence of interference, the receiver counts the number of chirps that are corrupted during the frame transmission. %

\begin{figure}[t]
    \centering
    \includegraphics[width=0.7\columnwidth]{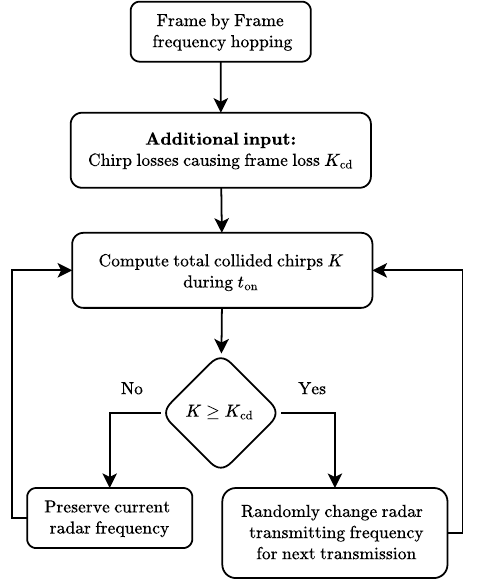}
    \caption{Frame-by-frame collision detection flow diagram.}
    \label{fig:flow_cd}
\end{figure}

The behavior of the collision detection algorithm is detailed in the flow diagram illustrated in Fig. \ref{fig:flow_cd}. %
The collision detection approach relies on the parameter $K_\text{cd}$, %
which %
represents the minimum number of corrupted chirps that triggers a change. %
More specifically, during the active transmission interval $t_\text{on}$, the total number of collided (i.e., corrupted) chirps $K$ is computed by the receiver of the given radar. %
If the number of collided chirps equals or exceeds the threshold $K_\text{cd}$ then the radar randomly changes its transmitting frequency. %

\begin{figure}[t]
\centering
\includegraphics[width=0.8\columnwidth]{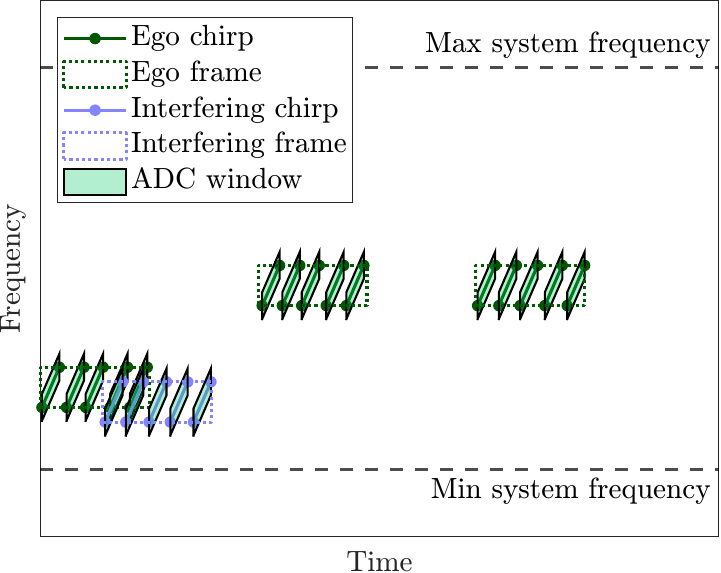}
\caption{Example of frame-by-frame collision detection mitigation method. }
\label{fig:fbf_cd_example}
\vspace{-\baselineskip}
\end{figure}

Figure~\ref{fig:fbf_cd_example} shows an example of the collision detection behavior obtained from the radar network simulator, with $K_\text{cd} = 2$. %
In this example, after the first transmission, the given radar changes its frequency, because two chirp collisions are detected, i.e., the collided chirps are equal to $K_\text{cd}$. %
In contrast, during the second transmission all chirps are received without collisions and the radar preserves the same starting frequency.

{%
\begin{table}[t]
\caption{Symbols and settings. %
}%
\footnotesize %
\centering
\begin{tabular}{p{2.8cm}p{1.8cm}p{1.3cm}p{1.3cm}}\hline\hline
\textbf{Parameter}  & \textbf{Symbol} & \textbf{Front} & \textbf{Corner} \\ 
\hline
\hline
\multicolumn{3}{l}{\textbf{Transceiver settings}}
\\
EIRP & $\EIRPdBm$ & 35 dBm & 15 dBm \\
Rx antenna gain & $\Gr$ & 30 dBi & 23 dBi\\
Field of view & FoV & 30° & 60° \\
Noise figure & $\NF$ & 15 dB & 15 dB\\
Carrier frequency & $\fzero$ & 140 GHz & 140 GHz\\
\hline
\multicolumn{3}{l}{\textbf{Signal settings}}
\\
Duration of a chirp & $\Tchirp$ & 5.14 $\mu$s & 10.3 $\mu$s\\
Chirp repetition time& $T_\text{rch}$ & 6.42 $\mu$s & 12.8 $\mu$s\\
\# of chirps in a frame & $\nch$ & 2000 & 1555 \\
Duration of a frame & $T_\text{f}$ & 12.8~ms & 20~ms \\
Duty cycle & $\duty$ & 0.5  & 0.25  \\
Frame repetition time %
& $\Tframe$ & 25.6~ms & 80~ms\\
Available bandwidth & $\BTOT$ & Variable & Variable \\
Chirp bandwidth & $B_\text{ch}$ & 150 MHz & 1.5~GHz \\
Max. beat frequency  & $\hat{f}_\text{b}$ & 68.1 MHz & 97.29~MHz \\
ADC bandwidth & $\BADC$ & 100 MHz & 100 MHz \\
ADC sample frequency & $2 \cdot \hat{f}_\text{b}$ & 136.2 MHz & 194.6 MHz \\
\hline
\multicolumn{3}{l}{\textbf{Radar metrics}}\\
Unambiguous range & $R_\text{max}=\frac{c\hat{f}_\text{b}T_\text{ch}}{2B_\text{ch}}$ & 350 m & 100 m \\
Range resolution & $\Delta R = \frac{c}{2B_\text{ch}}$ & 1.00 m & 0.10 m \\
Maximum velocity & $v_\text{max} = \frac{c}{4f_{0}T_\text{rch}}$ & 83.44 m/s & 41.85 m/s \\
Velocity resolution & $\Delta v = \frac{c}{2f_\text{0}T_\text{f}}$ & 0.0834 m/s & 0.0538 m/s \\
\hline
\multicolumn{3}{l}{\textbf{Reflection modeling}}
\\
Radar cross section & $\sigma$ & 10~$\text{m}^2$ & 10~$\text{m}^2$ \\
Minimum INR & $\INRmin$ & 0 dB & 0 dB \\
Maximum path length & $\dmax$ & 2694 m & 120 m \\
\hline 
\multicolumn{3}{l}{\textbf{Success/failure modeling}}
\\
\makecell[l]{Minimum chirp overlap} & $\xchirp$ & 0.5 & 0.5 \\[0.25mm]
\makecell[l]{Corrupted chirps\\for frame loss}
& $\kch$
& \makecell[l]{400\\[-1pt]\scriptsize (20\% of $\nch$)}
& \makecell[l]{311\\[-1pt]\scriptsize (20\% of $\nch$)}
\\[-2.5mm]
\makecell*[l]{Consecutive frame losses\\for system failure}
& $\Mf$ & 3 & 3 \\[-1mm]
\hline
\hline
\end{tabular}
\label{tab:settings}
\end{table}
}

\section{Performance Evaluation}

This section presents the results with reference to a dense highway scenario. The main settings, discussed hereafter, are also summarized in Table~\ref{tab:settings}.

\subsection{Simulation settings}\label{sec:W7_WiLab_simulator}

The results are obtained using a dynamic event-driven simulator written in Matlab, with vehicle positions obtained from the traffic simulator SUMO \cite{SUMO2018} and the potential interferers obtained from the WiLabVIsim \footnote{The tool is available at \href{https://github.com/V2Xgithub/WiLabVIsim}{https://github.com/V2Xgithub/WiLabVIsim}.} \cite{miccoli2021tool}. 

\subsubsection{Scenario} The scenario is a highway with 3 lanes per direction and 150 average vehicles per km, with positions and mobility reproduced by SUMO. This density represents a reference scenario with congested traffic to study a critical case from the radar interference perspective \cite{miccoli2021tool}. %

\begin{figure}[t]
\centering
\includegraphics[width=0.7\linewidth]{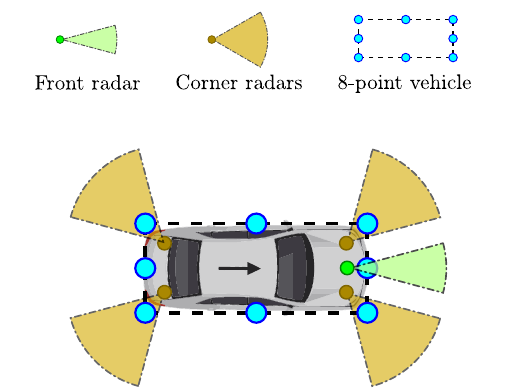}
\caption{Modeling of the vehicle, with either one front radar or four corner radars, and with eight points of reflection for the calculation of reflected interference.}
\label{fig:vehicleRadars}
\end{figure}

All vehicles are equipped with one long-range front radar and four mid-range corner radars, as shown in Fig.~\ref{fig:vehicleRadars}; front and corner radars use separate bands. For the evaluation of reflected interferers, the vehicles are treated as obstacles modeled using the eight reflecting points in Fig.~\ref{fig:vehicleRadars}, as in \cite{6202527}. The \ac{RCS} for reflection is assumed equal to $\sigma=10$~m$^2$, following the ETSI indications in \cite{ETSI_TR_103_593}. 
The signal parameters are taken from \cite{ETSI_TR_104_054} and reported in Table~\ref{tab:settings}. The adopted configuration targets operation around 140~GHz, as higher frequencies are being actively explored for future systems \cite{eu_decision_2026_883}. The observed trends are expected to generalize across frequency bands.

\subsubsection{Benchmarks}
We compare our proposals with the following benchmarks \cite{11422277}:

\begin{itemize}
\item \textit{Baseline}: The starting frequency of each radar is randomly selected between $\fstartTOT$ and $\fstopTOT-\BW$ according to a uniform distribution. The same starting frequency is kept by the given radar and never changed.

\item \textit{Frame-by-frame (FbF)\acused{FbF} frequency hopping}: The starting frequency of the first chirp of each new frame is randomly selected within $\fstartTOT$ and $\fstopTOT-\BW$ with uniform distribution, and then the same frequency is used for the following chirps of the same frame. 

\item \textit{Chirp-by-chirp (CbC)\acused{CbC} frequency hopping}: The starting frequency of each chirp is randomly selected within $\fstartTOT$ and $\fstopTOT-\BW$ with uniform distribution.

\end{itemize}

\bigskip
The baseline approach is likely representative of current market radar implementations, and the chirp-by-chirp provides a relevant benchmark while being more theoretical, since changing the frequency from chirp to chirp arbitrarily is still challenging to implement in practice.

\subsubsection{Output Metric}\label{sec:outputmetric}

The performance of the proposed techniques is evaluated in terms of the average time between two system failures, calculated as
\begin{equation}\label{eq:t_fail}
    \Tfail = \Tframe \cdot \frac{1}{P_\text{fail}}
\end{equation}
where $P_\text{fail}$ is the probability of having a system failure, which corresponds to the loss of $\Mf=3$ consecutive frames, where each loss is caused by at least 20\% of corrupted chirps. This probability is, in turn, estimated by the ratio between the number of system failures experienced by each radar and the maximum possible number of failures that would occur if all transmissions were lost.

\subsection{Results}\label{subsec:Results}

\subsubsection{Preliminary Considerations for Collision Avoidance}\label{sec:ca_considerations}

\begin{figure}[t]
    \centering
    \subfloat[]{%
    \includegraphics[width=0.45\textwidth]{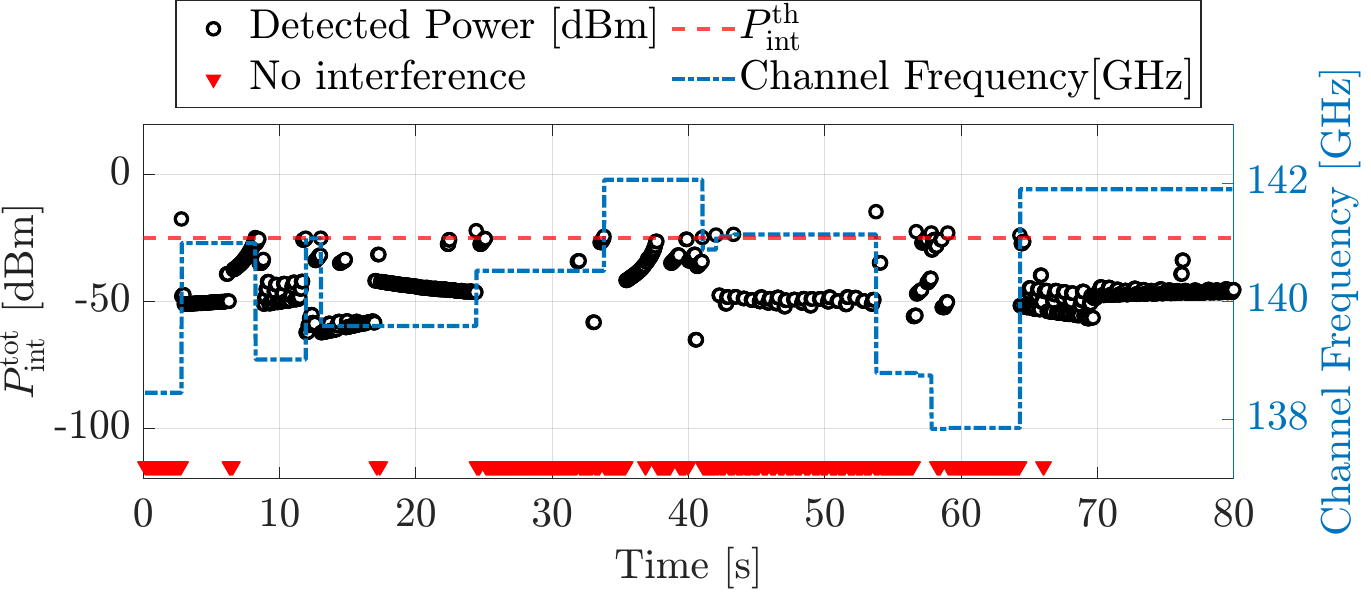}
    \label{fig:ca_interference_vs_freq_-25dbm}
}
\hfill
\subfloat[]{%
    \includegraphics[width=0.45\textwidth]{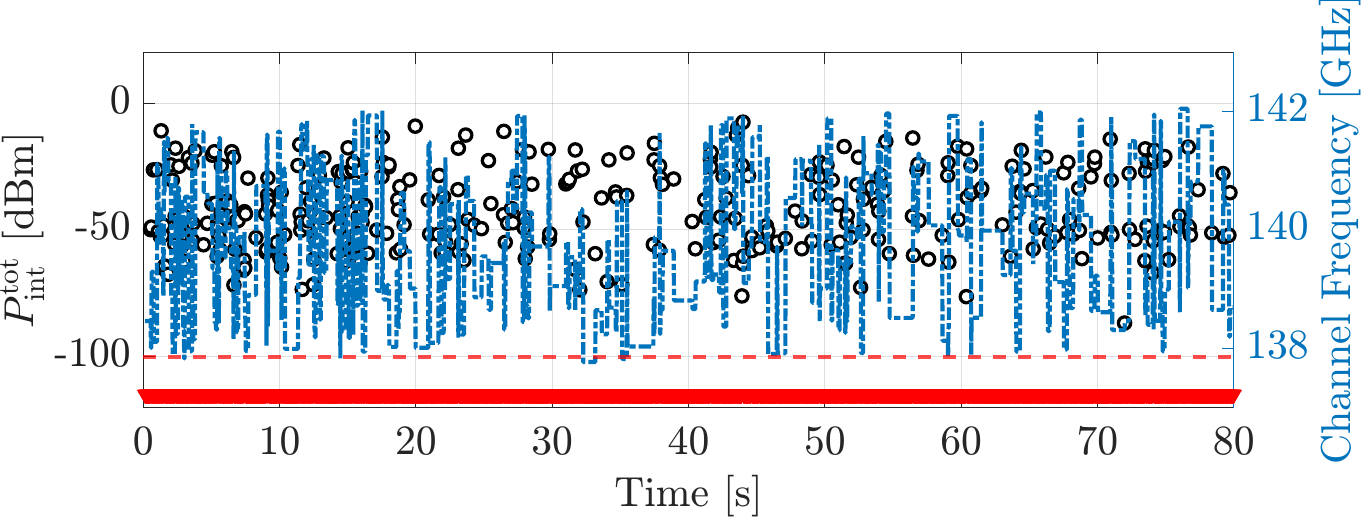}
    \label{fig:ca_interference_vs_freq_-100dbm}
}
    \caption{Time evolution of the detected interference power and selected transmission frequency for a given radar using the FbF-CA technique; (a) with threshold $P_\text{int}^\text{th}=-25$ dBm; (b) with threshold $P_\text{int}^\text{th}=-100$ dBm.
    }  
    \vspace{-\baselineskip}
    \label{fig:ca_interference_vs_freq}
\end{figure}

Figure \ref{fig:ca_interference_vs_freq} provides preliminary results on the \ac{FbF-CA} technique, with focus on the detected interference power threshold $P_\text{int}^\text{th}$. Front radars are here assumed, but similar conclusions were derived with corner radars.

Figures~\ref{fig:ca_interference_vs_freq_-25dbm} and \ref{fig:ca_interference_vs_freq_-100dbm}, in particular, illustrate the time evolution of the detected interference power and the corresponding channel selection for a generic radar and for two different values of $P_\text{int}^\text{th}$. The left y-axis reports the detected interference power, while the right y-axis indicates the selected transmission frequency. The red dashed horizontal line represents the interference threshold, black circles denote the instantaneous detected interference power, and red triangles indicate the detection of a free channel.
Figure~\ref{fig:ca_interference_vs_freq_-25dbm} refers to a high threshold value, i.e., $-25$~dBm. In this case, the detected interference power exceeds the threshold only occasionally and therefore the radar attempts to mitigate interference by changing its transmission frequency only when the interfering power is very high. 
Figure~\ref{fig:ca_interference_vs_freq_-100dbm} shows instead the same metrics assuming a low threshold value, i.e., $-100$~dBm. Under this condition, the detected interference power exceeds the threshold more frequently, leading to more frequent changes in the radar starting frequency. 
Notably, by comparing Figs. \ref{fig:ca_interference_vs_freq_-25dbm} and \ref{fig:ca_interference_vs_freq_-100dbm}, it can be observed that a lower threshold leads to more frequent frequency hopping events, %
which in turn increases the likelihood of identifying free channels, as indicated by the larger number of red triangles in Fig.~\ref{fig:ca_interference_vs_freq_-100dbm}. More experiments, not shown here for brevity, suggest to keep this threshold as low as possible.

\subsubsection{Preliminary Considerations for Collision Detection}

\begin{figure}[t]
\centering
\includegraphics[width=0.8\columnwidth]{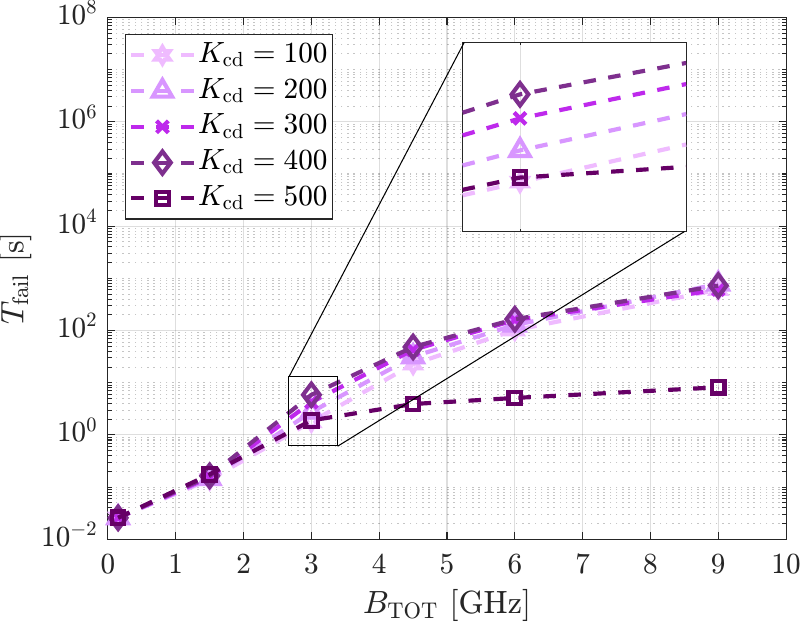}
\caption{Average time between system failure using the FbF-CD technique with different $K_\text{cd}$; front radars, 150 vehicles/km.}
\vspace{-0.5cm}
\label{fig:cd_varying_Kcd}
\end{figure}

With reference to the \ac{FbF-CD} technique, a critical parameter is the threshold on the number of corrupted chirps, $K_\text{cd}$. %
Also in this case, the analysis shown in Fig.~\ref{fig:cd_varying_Kcd} refers only to front radars, but similar conclusions were drawn for corner radars.

In particular, Fig.~\ref{fig:cd_varying_Kcd} illustrates the system performance in terms of average time between system failures as a function of the total system bandwidth for various values of $K_\text{cd}$. As can be observed, when the collision detection parameter satisfies $K_\text{cd} \leq K_\text{ch}$, the performance remains nearly unchanged. Among these cases, the configuration $K_\text{cd} = K_\text{ch}$, corresponding to the purple curve with diamond markers, exhibits a slightly better behavior. In contrast, when $K_\text{cd}$ exceeds  $K_\text{ch}$, the performance degrades significantly, as shown by the square-marked curve. A zoomed view further highlights that the $K_\text{cd} = K_\text{ch}$ condition achieves slightly better performance than the other cases.

Although this result is partly expected, since the optimal setting is consistent with the definition of a frame loss, an important observation is that the choice of $K_\text{cd}$ does not appear critical, provided that it is not too conservative in triggering frequency changes.

\subsubsection{Performance comparison}
\begin{figure}[t]
\centering
\includegraphics[width=0.8\columnwidth]{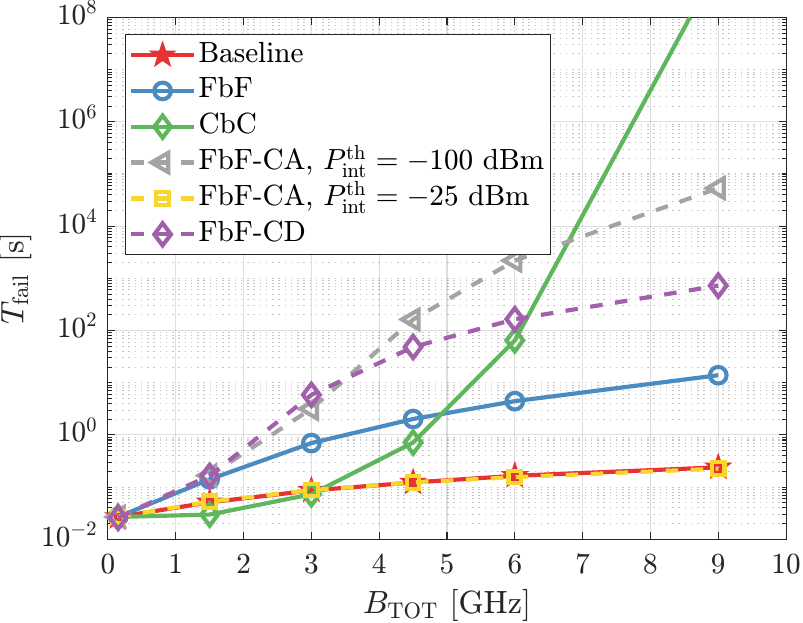}
\caption{Average time between system failure vs. available bandwidth; front radars, 150 vehicles/km. }
\label{fig:TF_front_150v}
\end{figure}

\begin{figure}[t]
\centering
\includegraphics[width=0.8\columnwidth]{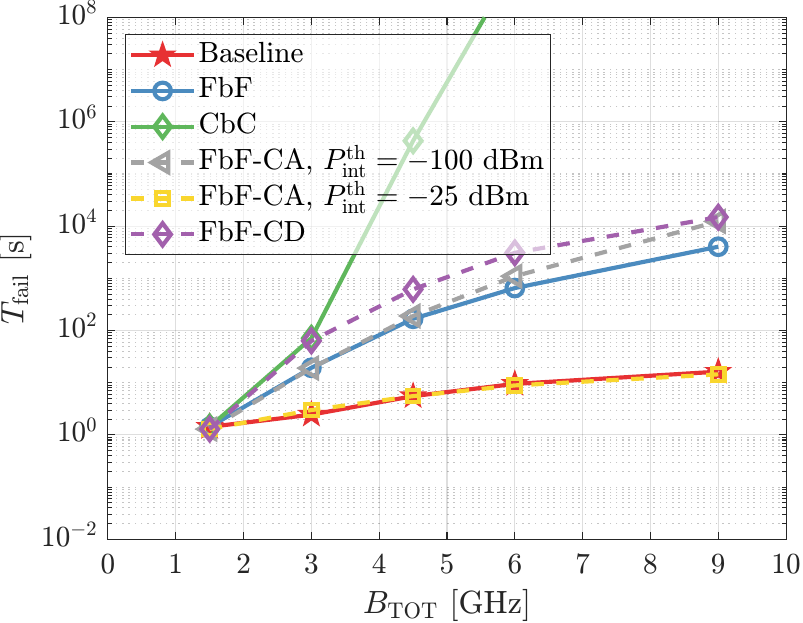}
\caption{Average time between system failure vs. available bandwidth; corner radars, 150 vehicles/km.}
\label{fig:TF_corner_150v}
\vspace{-\baselineskip}
\end{figure}

In Figs.~\ref{fig:TF_front_150v} and \ref{fig:TF_corner_150v}, \ac{FbF-CA} and \ac{FbF-CD} are compared with the baseline, \ac{FbF}, and \ac{CbC}.
Figure~\ref{fig:TF_front_150v}, in particular, shows the results in terms of average time between system failures versus the system bandwidth, assuming front radars. Looking first at the baseline, \ac{FbF}, and \ac{CbC} techniques (the three solid lines), an improvement of the performance can be observed as the system bandwidth increases. Among these, the \ac{CbC} technique (green solid curve with diamond markers) represents the worst solution for small system bandwidth, while becoming the best performing approach when the total system bandwidth is large. 

Focusing on the \ac{FbF-CA} approach, Fig. \ref{fig:TF_front_150v} reports two configurations: a high detected interference threshold of $-25$ dBm (yellow dashed curve with square markers) and a low threshold of $-100$ dBm (grey dashed curve with triangle markers). When a high threshold is adopted, radars tend to preserve the same transmitting frequency over time, resulting in behavior similar to the baseline case where each radar maintains a fixed starting transmission frequency. Conversely, decreasing the interference threshold enables more frequency changes and this increases the probability of detecting free channels. Additional experiments, not shown here for brevity, demonstrated that using intermediate values for the threshold lead to intermediate results. Comparing the \ac{FbF-CA} with the other methods, the results show that \ac{FbF-CA} with a low threshold outperforms \ac{FbF}, since collision avoidance triggers frequency changes only when required, leading to a more efficient use of the spectrum. \ac{CbC} remains preferable only if the bandwidth is very large.

Figure \ref{fig:TF_front_150v} also reports the results obtained with \ac{FbF-CD} technique, with $K_\text{cd}$ set to $20\%$ of the chirps in a frame. 
Collision detection (purple dashed curve with diamond markers) represents the best technique for low system bandwidth, as it enables frequency changes only when strictly required. However, as the system bandwidth increases, \ac{FbF-CD} is outperformed by \ac{FbF-CA}. 
This behavior can be attributed to the ability of collision avoidance to proactively detect the presence of interference and trigger a prompt frequency change.
With increasing bandwidth, collision avoidance allows radars to change frequency even in the presence of low interference levels, whereas collision detection triggers frequency changes only after a frame loss has already occurred. %

Figure~\ref{fig:TF_corner_150v} finally illustrates the same comparison for corner radars. This case implies a larger bandwidth of the chirp, thus a smaller relative available bandwidth, and a lower number of potential interferers. In this case, the FbF-CA improves the performance compared to FbF only when the available bandwidth exceeds 6 GHz, and FbF-CD outperforms FbF-CA except when 9 GHz is available. %
Furthermore, in this case, neither of these techniques approaches the performance achieved by the \ac{CbC} technique, which, however, is difficult to implement in practice, as already stated. %

\section{Conclusion}\label{Sec:conclusion}

In this work, we assessed the effectiveness of two proposed interference mitigation techniques based on measurements performed during idle intervals or during transmissions, called \ac{FbF-CA} and \ac{FbF-CD}, respectively. Assuming correlated interference and realistic dense highway scenarios, the results demonstrate that the two approaches significantly reduce the impact of interference compared to both a baseline solution where radars do not change the frequency configuration and a solution where the carrier frequency randomly varies at every  burst of chirps (called frame). When the number of interferers is large and the available bandwidth is limited, the two solutions also proved to outperform a more ideal solution where the frequency changes at every chirp. %
Future work should extend the analysis to heterogeneous radar scenarios, to evaluate the robustness of the proposed techniques under radars with mixed configuration or uncorrelated interference conditions.

\bibliographystyle{IEEEtran}
\bibliography{biblio}

\end{document}